\documentclass[twocolumn,showpacs,preprintnumbers,superscriptaddress,amsmath,amssymb]{revtex4}

\usepackage{graphicx}
\usepackage{dcolumn}
\usepackage{bm}

\begin{document}

\preprint{APS/}

\title{Theory of the electrical transport in tilted layered superconductive\\
Josephson junctions}

\author{C. Nappi}
\email{c.nappi@cib.na.cnr.it}
\affiliation{%
CNR Istituto di Cibernetica ''E. Caianiello''
I-80078, Pozzuoli, Napoli, Italy}%

\author{S. De Nicola}
\affiliation{%
CNR Istituto Nazionale di Ottica I-80078, Pozzuoli, Napoli, Italy and INFN sezione di Napoli, I-80126 Napoli, Italy}%

\author{M. Adamo}
\affiliation{%
CNR Istituto di Cibernetica ''E. Caianiello''
I-80078, Pozzuoli, Napoli, Italy}%

\author{E. Sarnelli}
\email{e.sarnelli@cib.na.cnr.it}
\affiliation{%
CNR Istituto di Cibernetica ''E. Caianiello''
I-80078, Pozzuoli, Napoli, Italy}%

\date{\today}

\begin{abstract}
We present a theory of the Josephson effect in a twofold-tilted
Josephson junction made by d-wave anisotropic layered superconductors. We
find the appearance of an intrinsic electrical resistance that
arises from the misalignment of the superconductive planes (the
$CuO_2$ planes in YBCO) in the two electrodes. This intrinsic
contribution to the tunnel barrier has several non-trivial
consequences. The result is relevant for understanding the electric
transport properties of [100] tilt and [100] tilt-tilt Josephson
junctions based on d-wave superconductors.
\end{abstract}

\maketitle

 Grain boundary Josephson junctions (GBJs) have been fabricated and studied immediately after the discovery
 of High Tc superconductors\cite{Dimos0}. Nonetheless the mechanism of the supercurrent transport in these devices is
 still  matter of scientific investigation and argument of general physical
 interest \cite{Nature, Kupriyanov}.
 It is well known that the dc Josephson effect can be described by means of a
 scattering theory using the Bogoliubov-de Gennes (BdG) equations \cite{degennes}.
Characteristic localized states of quasiparticles, known as Andreev
bound states, are found in the superconductive energy gap region
through which the Josephson supercurrent flows between the two
electrodes \cite{furusaki}.
This kind of theoretical approach has
been successfully used in describing analytically the physics of in-plane grain
boundary  cuprate Josephson junctions ([001] tilt) where a
two-dimensional scattering theory \cite{tanaka96,tanaka, Lofwander} finely
incorporates both the propagation of the quasiparticles along the
Cu-O planes and the anisotropic d-wave-induced symmetry of the pair
potential \cite{Hil-Mann}. Moreover in ref \cite{Nature} a
numerical approach, also based on the BdG equations,
has enriched the understanding  of the role of charge inhomogeneities in limiting supercurrents in
[001] GBJs.

\begin{figure}
\includegraphics[width=22pc]{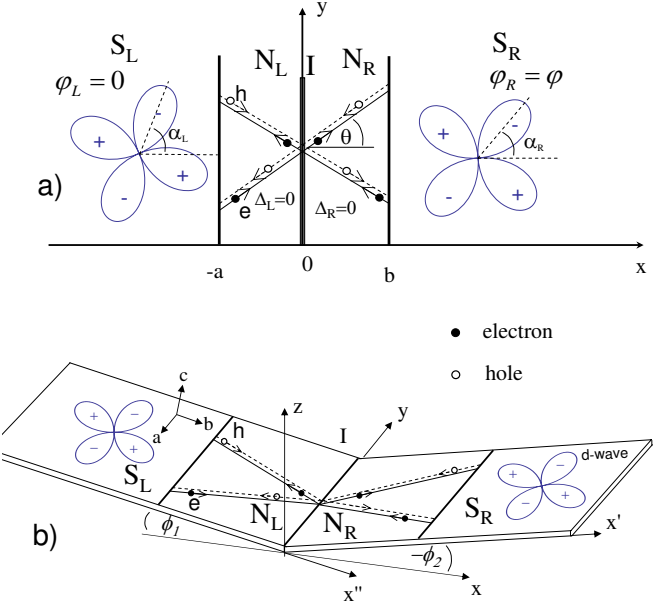}
\caption{\label{fig:SNINS} a) z axis view of a layered SNINS [100]
tilt-tilt junction. The thickness of the normal region is a+b, the
d-wave superconducting regions occupy the space defined by $x < -a$
and $x > b$. The insulating barrier (I) is located at x = 0. b) 3d
view of the conducting planes with the indicated tilt angles
$\phi_1$ and $\phi_2$. The d-symmetry pair potential is represented
by $\Delta^L(\theta) = \Delta_0 \cos(2\theta-2\alpha_L);
\Delta^R(\theta) = e^{i \varphi}\Delta_0 \cos(2\theta-2\alpha_R)$,
 $\varphi=\varphi_R-\varphi_L$ is the global phase difference
between the two superconducting regions.  The solid and dashed lines
represent the electron-like and the hole-like elementary excitation
trajectories extending over the two planes, respectively.}
\end{figure}

 However,
a number of experiments  have been carried
out on [100] YBCO  GBJs\cite{Dimos_PRB,Carillo,Divin0},  where the Cu-O planes of one or both electrodes are
 tilted by an arbitrary angle ($\phi_1$ and $\phi_2$ in Fig. \ref{fig:SNINS}) with respect the
substrate surface,  implying that the relevant geometry
of the junction is no longer in-plane.
 In particular, relatively high values
of the $I_cR_n$ products \cite{Divin0,Sarnelli,Divin}  have been observed in such
YBCO GBJs.
 [100] junctions seem to be very
promising for
developing new sensors of radiation in the terahertz band
\cite{Divin_Poppe, Stepantsov} and observing
macroscopic quantum properties \cite{bauch}. Indeed Kawabata {\it et al} \cite{Shiro} have
pointed out that macroscopic quantum effects in YBCO devices may be better
observed in junctions showing electrodes with  pair
potential lobes aligned each other ($d_0/d_0$ junctions). This
configuration can be obtained with out-of-plane [100] GBJs
geometries, whereas [001] GBJs are not suitable for fabricating
$d_0/d_0$ junctions \cite{Hil-Mann}.

In spite of all these circumstances, as of today, and differently from
 the [001] tilt junction case, a specific theoretical
treatment of the transport properties in [100] junctions is lacking \cite{Golubov}.
One can probably individuate
emblematically the obstacle of developing a model theory for
this case in
 the non-trivial modification suffered by the quasiclassical trajectories of the excitations.
 Indeed such trajectories are no longer confined to a single
plane as in [001] layered
 junctions \cite{Nature,tanaka96,tanaka,Lofwander}, but lay in planes tilted with respect to the substrate
  plane by the misorientation angles $\phi_1, \phi_2$, see Fig.\ref{fig:SNINS}.

 In this Letter we take an important step
towards the effective evaluation of the peculiar transport
properties of YBCO [100] tilt junction ($\alpha_L$=$\alpha_R=0$ in Fig. \ref{fig:SNINS}) and [100] tilt-tilt junctions
(junctions in which one or both angles $\alpha_L$, $\alpha_R$ are different from zero).

We
focus on the dc Josephson effect and find that the influence of the different tilting of the
conduction planes with respect to the barrier plane (on passing from
one electrode to the others) brings about an enhanced normal state
electric resistance of the junction. The key result is an analytical expression for the Andreev
spectrum for the quasiparticles, which retains the influence of the tilting of the Cu-O
planes as well as the effects of
the d-wave anisotropic symmetry of the pair potential.

 In order to describe charge transport in a mesoscopic [100] YBCO grain boundary
 Josephson junction, we assume a
 superconductor-normal-insulator-normal-superconductor (SNINS) model
 \cite{Lofwander,navacerrada,Testa,Herrera}. The superconductive material  considered
 is a d-wave anisotropic layered cuprate.  The presence of
normal regions of the order $\sim\xi_0=\hbar v_F/\pi\Delta_0$, the ballistic coherence length \cite{Lofwander-Thesis},
 is introduced for modeling the mechanism of the Andreev reflection and a possible suppression of the order parameter near the junction.

Furthermore we suppose that the
quasiparticles are constrained to move exclusively along the Cu-O
planes of the two electrodes. This last assumption is based on the fact that the normal conductivity
in c axis direction in YBCO is about one hundred times smaller than that along the $ab$ plane. In the superconducting state
this strong anisotropy persists in the supercurrent distribution owing to a
large ratio  $\lambda_{c}^2/\lambda_{ab}^2$ where
$\lambda_{c}$ ($\lambda_{ab}$) is the London penetration depth across the planes (in the planes)
 \cite{Kupriyanov, Friedmann}. \\
 The junction barrier, the plane $x=0$
in the frame xyz indicated in Fig.\ref{fig:SNINS}, is normal to the
substrate and is considered perfectly flat.  This reflects the fact that in a 'valley' type morphology junction,
like the one represented in Fig. \ref{fig:SNINS}b,  there is
almost perfect matching of the conducting planes at the GBJ \cite{Ogawa}.
These latter are tilted with respect the
substrate plane $z=0$, by the angles $\phi_1$ and $-\phi_2$
 in the left ($x<0$) and in
the right ($x>0$) electrode, respectively. The two $SN$ interfaces
 are normal to the x
axis, as indicated in Fig. \ref{fig:SNINS}. The coupled
motion of hole-like ($v$) and electron-like ($u$) components of the
wave function $\Psi$ in the Cu-O planes is described by the 2d
quasiclassical Bogoliubov de Gennes  equations \cite{Bruder}. For the
plane $x'-y$ and $x>0$ they write (analogous equations hold in the
plane $x''-y$ for $x<0$)
\begin{equation}
\label{eq:J_1}  \left\{
\begin{array}{rl}
Eu(x',y)=\hat{h}(x',y)u(x',y)+\Delta_R(\hat k) \Theta(x'-b) v(x',y)\\
\\
Ev(x',y)=-\hat{h}(x',y)v(x',y)+\Delta_R(\hat k)\Theta(x'-b) u(x',y)
\end{array} \right.
\end{equation}
where
$\hat{h}(x',y)=-\hbar^2(\partial^2/\partial x'^2+
\partial^2/\partial y^2)/{2 m}+V(x)-E_F .$
$\Theta(z)$ is the Heaviside step function and $E_F=\hbar^2
k_F^2/2m$ is the Fermi energy. The insulating barrier is modeled by
a delta function, $V(x) = U_0\delta(x)$, $U_0$ being the Hartree
potential. The spatial dependence of the pair potential is assumed to have
a step-functional form described by $\Delta(\hat{k},\textbf{r})=
\Theta(-a-x)\Delta_L(\hat{k})+\Theta(x-b)\Delta_R(\hat{k})$.  The
bound states solutions of Eqs (\ref{eq:J_1})  for $\Psi(x')$
and $\Psi(x'')$ [$\Psi(x',y) = \Psi(x')\exp{(ik_yy)},   \Psi(x'',y)
=\Psi(x'')\exp{(ik_yy)}]$ in the $S_{L}$, $N_{L}$, $N_{R}$ and
$S_{R}$ regions, including normal and Andreev reflections of holes
and electrons, are respectively
\begin{eqnarray}
\label{eq:PSISd}   \Psi^{S_{L}}(x'') =   a_B(E)\left(
\begin{array}{rl}
u_0^{L,-} \nonumber \\
 v_0^{L,-}e^{-i \varphi_-^L}
\end{array}
\right)e^ {-i k^{L,e}_{-} x''}+ \nonumber \\ b_B(E)\left(
\begin{array}{rl}
v_0^{L,+} \nonumber \\
 u_0^{L,+}e^{-i \varphi_+^L}
\end{array}
\right)e^ {i k^{L,h}_{+} x''}
 \nonumber \\
\Psi^{N_L}(x'') =  \left(
\begin{array}{rl}
1\\
0
\end{array}
\right)\left(U_1e^{ik_1^ex''}+U_2e^{-ik_1^e x''}\right) +\nonumber
\\ \left(
\begin{array}{rl}
0\\
1
\end{array}
\right)\left(V_1e^{ik_1^h x''}+V_2e^{-ik_1^h x''}\right) \nonumber
\\
\Psi^{N_R}(x') =  \left(
\begin{array}{rl}
1\\
0
\end{array}
\right)\left(U_3e^{ik_1^ex'}+U_4e^{-ik_1^e x'}\right)+ \nonumber \\
\left(
\begin{array}{rl}
0\\
1
\end{array}
\right)\left(V_3e^{ik_1^h x'}+V_4e^{-ik_1^h x'}\right) \nonumber \\
 \Psi^{S_{R}}(x') =   c_B(E)\left(
\begin{array}{rl}
u_0^{R,+} \nonumber \\
 v_0^{R,+}e^{-i (\varphi_+^R+\varphi)}
\end{array}
\right)e^ {i k^{R,e}_{+} x'}+ \nonumber \\ d_B(E)\left(
\begin{array}{rl}
v_0^{R,-}  \\
 u_0^{R,-}e^{-i (\varphi_-^R+\varphi)}
\end{array}
\right)e^ {-i k^{R,h}_{-} x'}
\end{eqnarray}
with
\begin{eqnarray}
&&u_0^{\beta,\pm}=\left[\frac{1}{2}\left(1+i\frac{\Omega^\beta_\pm}{E}\right)\right]^{1/2},
v_0^{\beta,\pm}=\left[\frac{1}{2}\left(1-i\frac{\Omega^\beta_\pm}{E}\right)\right]^{1/2}\nonumber
\\ && k^{\beta,e}_\pm =
\left(k_F^2-k_y^2+i\frac{2m\Omega^{\beta}_\pm}{\hslash^2}\right)^{1/2},\nonumber \\
&&k^{\beta,h}_\pm =
\left(k_F^2-k_y^2-i\frac{2m\Omega^{\beta}_\pm}{\hslash^2}\right)^{1/2}, \Omega^\beta_\pm=\left(|\Delta^\beta_\pm|^2-E^2\right)^{1/2} \nonumber \\
&&k_1^e=\left(k_F^2-k_y^2+\frac{2mE}{\hslash^2}\right)^{1/2},k_1^h=\left(k_F^2-k_y^2-\frac{2mE}{\hslash^2}\right)^{1/2}
\end{eqnarray}
$\beta$ denoting right ($R$) or left ($L$), $k^{\beta,e}_\pm$, $k^{\beta,h}_\pm$ are
 the wave numbers of the electron-like and hole-like quasiparticles respectively,
  see Fig.\ref{fig:SNINS}a, that move in the superconducting regions,
  $k_1^e$, $k_1^h$ the wave numbers of the electrons and holes in the normal regions respectively, and
$E\leq|\Delta^\beta_\pm|$. The conducting layers in the
electrodes are treated as planes parallel to
the $x''-y$ plane or to the $x'-y$ plane. The propagation
coordinates $x'$, $x''$ in the left and right electrode are given by
$x'=x \cos \phi_1 +z \sin \phi_1, x''=x \cos \phi_2 -z \sin \phi_2$,
respectively. For d-wave symmetry the effective pair potential is
modeled as $\Delta_{\pm}^L=\Delta_0|\cos(2(\theta \mp \alpha_L))|
e^{i \varphi_{\pm}^L}, \Delta_{\pm}^R=\Delta_0 |\cos(2(\theta \mp
\alpha_R))|e^{i (\varphi+\varphi_{\pm}^R)}$
\cite{tanaka96,tanaka}, where $\varphi_{\pm}^\beta$ are the
phases as felt in a normal reflection by the electrons (+) and holes
(-), $\varphi$ is the global
 phase difference between the two superconducting regions and $\alpha_L$, $\alpha_R$
 are the angles between the crystallographic a-axis of the left and right superconductors  and the
  normal to the y axis. $\theta$ is the incident
  angle of the quasiparticle trajectories to the y axis.
  Boundary conditions have to be specified at the three different
interfaces, $x=-a$, $x=0$ and $x=b$, to determine the twelve unknown
coefficients in $\Psi(x')$ and $\Psi(x'')$.  These conditions are:
the continuity of the function $\Psi$
 and its derivatives, $d\Psi/dx'$ and $d\Psi/dx''$, across the clean interfaces $x=-a,x=b$, respectively.
 Moreover, at the normal-insulator-normal
interface (plane $x=0$) where the change in the propagation
direction from $x''$ to $x'$ occurs, boundary conditions are the
continuity of $\Psi$ and
 the discontinuity of $d\Psi/dx$ imposed by the presence of the $\delta$-function barrier insulator.
Then, since $d/dx=\cos \phi_1 d/dx'$ and $d/dx=\cos \phi_2 d/dx''$,
we write the matching conditions at $x = 0$ as
 \begin{eqnarray}\label{eq:vector_comp}
&& \Psi_{N_L}(0)=\Psi_{N_{R}}(0)  \\ &&
\cos{\phi_2}\frac{d\Psi_{N_{R}}(0)}{dx''}-\cos{\phi_1}\frac{d\Psi_{N_L}(0)}{dx'}=\frac{2mU_0}{\hslash
^2}\Psi_{N_L}(0) \nonumber
 \end{eqnarray}

 Using Eq. (\ref{eq:PSISd}) and imposing the above
conditions, under the assumption of perfect
retro-reflectivity ($E_F\gg E, |\Delta_\pm^\beta|$) of the Andreev
reflections \cite{kashiwaya2}, we derive an homogeneous linear system of
equations for the the twelve unknown coefficients. The  condition of
existence of solutions, in addition to the trivial one, provides the
following spectral equation:
\begin{widetext}
\begin{eqnarray} \label{eq:energy2}
\left(\Gamma_-^L \Gamma_-^R-e^{-i (\gamma_-+\varphi)} e^{-iDr}
\right) \left(\Gamma_+^L \Gamma_+^R-e^{i (\gamma_++\varphi)}
e^{-iDr} \right) +  Z^2_{eff} \left(\Gamma_-^L
\Gamma_+^L-e^{-i \gamma_L} e^{-2iAr} \right) \left(\Gamma_-^R
\Gamma_+^R-e^{i \gamma_R} e^{-2iBr} \right)=0
\end{eqnarray}
\end{widetext}
 with
\begin{eqnarray}\label{eq: definitions}
&&\Gamma_\pm^\beta=\frac{E-i\Omega^\beta_\pm}{|{\Delta^\beta_\pm}|},\gamma_{\beta}=\varphi^\beta_+
- \varphi^\beta_- ,
\gamma_{\pm}=\varphi^R_{\pm} - \varphi^L_{\pm} \nonumber \\
&&r=\frac{k_F}{\cos \theta} \frac{E}{E_F}, A=\frac{a}{\cos
\phi_2}, B=\frac{b}{\cos \phi_1},D=A+B
\end{eqnarray}
and where the effective barrier parameter $Z_{eff}$, defining the
electrical transparency of the junction $T_{eff}=1/(1+Z_{eff}^2)$),
appears
\begin{eqnarray}\label{eq:Zeff}
&&Z_{eff}(\theta,\phi_1,\phi_2)=\sqrt{\frac{4Z(\theta)^2+\left(\cos \phi_1-\cos
\phi_2\right)^2}{4 \cos \phi_1 \cos \phi_2}}
\end{eqnarray}
with $Z(\theta)={Z_0}/{\cos \theta}$ and $Z_0={k_F U_0}/{2 E_F}$ the
usual barrier  parameter. With the definition Eq. (\ref{eq:Zeff}) we may write the spectral equation
in the same form which holds for the [001] case described
in \cite{kashiwaya2} and \cite{Herrera}. However Eq. (\ref{eq:Zeff}) shows a new contribution to the barrier,
coming from the specific geometrical configuration depending on the angles $\phi_1$ and $\phi_2$.
Equation
(\ref{eq:energy2}) provides the quasiparticle energy levels as a
function of the superconducting phase difference $\varphi$ for the
case of layered SNINS d-wave [100] tilt-tilt junctions and other
limits (SIS, SNS, INS). When the junction is in the clean limit,
($Z_0=0$) the transparency of [100] junctions does not reach the
unitary value but it is still limited to
\begin{equation}
\label{eq:Teff_ballistic}
T_{eff}^0=\frac{4\cos\phi_1\cos\phi_2}{(\cos\phi_1+\cos\phi_2)^2}
\end{equation}
which is $\theta$-independent. Eq. (\ref{eq:Teff_ballistic}) embodies the pure effect on the
 junction transparency of the misalignment
 of the conducting planes between the two electrodes.
  The zero-temperature normal-state conductance $G_n$ is evidently also affected
  by the angles
$\phi_1$ and $\phi_2$ and can be written, in terms of the average transparency
 $\langle T_{eff}\rangle$, as
\begin{eqnarray}
\label{eq:GN2} &&G_n=2\frac{e^2k_F L_y}{h}\langle
T_{eff}\rangle \nonumber \\ && \langle T_{eff}\rangle=
T_{eff}^0\left[1-\frac{C^2 \coth^{-1}
\left(\sqrt{1+C^2}\right)}{\sqrt{1+C^2}} \right]
\end{eqnarray}
where $C=2 Z_0/ (\cos \phi_1+\cos\phi_2)$, and $2\langle T_{eff}
\rangle=\int_{-\pi/2}^{\pi/2}d\theta \cos (\theta)
T_{eff}(\theta,\phi_1,\phi_2)$. Therefore, the effect of the tilting
of the planes in SNINS junction manifests as {\it an excess of
electrical resistance in the normal state}.  This effect persists
also in the limit of zero size normal regions ($a=b=0$).

\begin{figure}
\includegraphics[width=22pc]{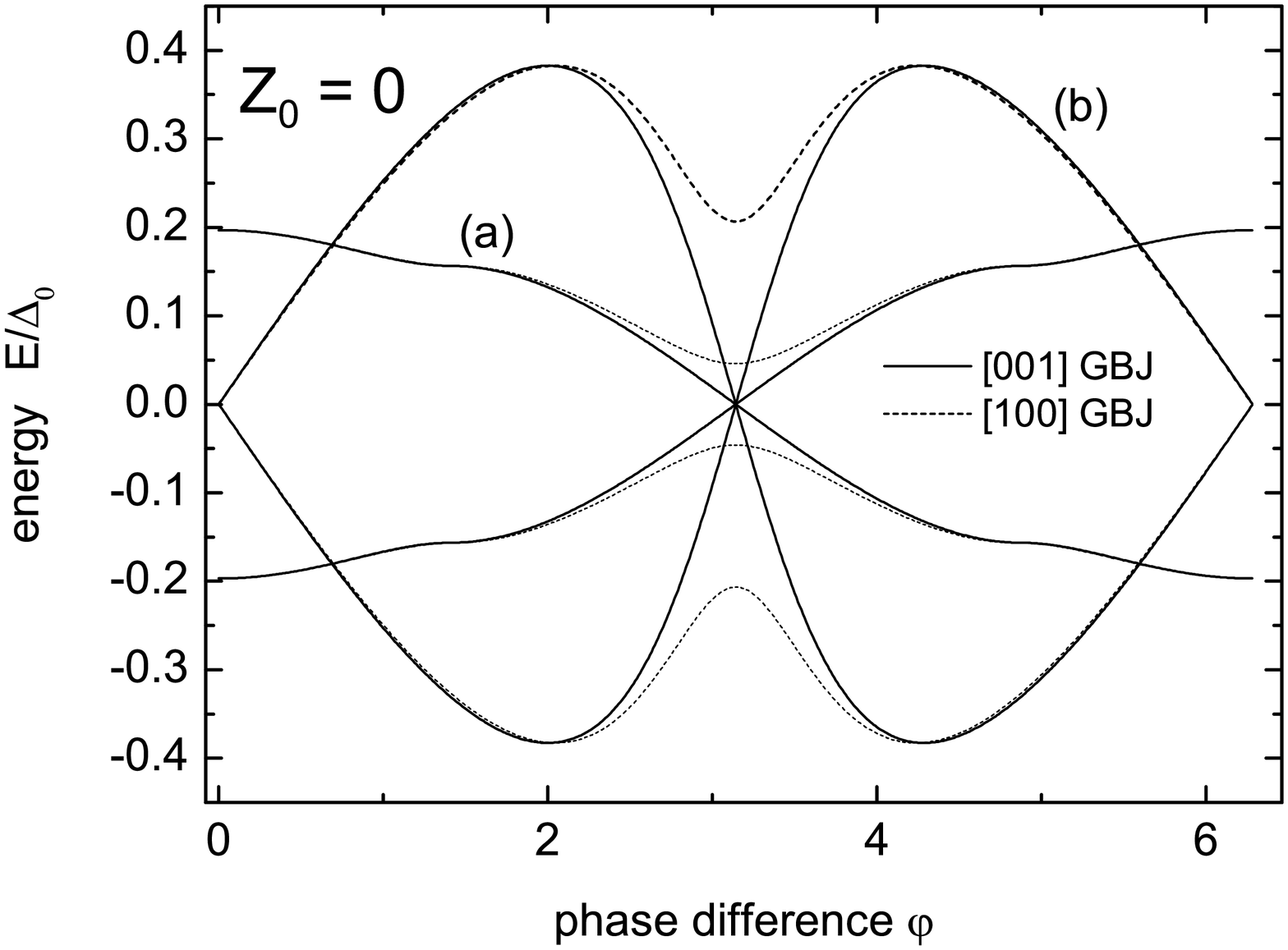}
\caption{\label{fig:specrum1} Comparison between Andreev level
spectra of d-wave SNS $45^\circ$ mirror
($\alpha_L=-\alpha_R=22.5^\circ$) grain boundary junctions
calculated in the clean limit ($Z_0=0$) for [001] and [100]
configurations, respectively. Solid and dashed curves refer to [001]
GBJs with $\phi_1=\phi_2=0$, and [100] GBJs ($\phi_1=0$ and
$\phi_2=45^\circ$) respectively. Curves a): edgegap-like state
Andreev levels for a trajectory with quasiparticle incidence angle
$\theta=\pi/10$; curves b) midgap-like state Andreev levels  with
$\theta=3\pi/16$.}
\end{figure}
\begin{figure}
\includegraphics[width=22pc]{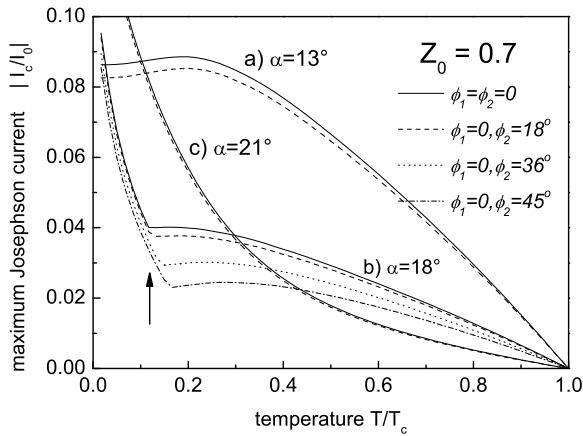}
\caption{\label{fig:josephson} Comparison of the temperature
dependence  of the absolute value of the Josephson critical current
in d-wave [001] tilt (solid curves) and [100] tilt-tilt mirror
($\alpha_L=-\alpha_R=\alpha$) grain boundary junctions with barrier
parameter $Z_0=0.7$ for three representative misorientation angles
$\alpha$. a) $\alpha=13^\circ$, b) $\alpha=18^\circ$, and c)
$\alpha=21^\circ$. For $\alpha=18^\circ$ arrow indicates the sign
inversion of the critical current ($0-\pi$ crossover) in [001] GBJs.
The remaining curves in b) show a shift of the crossover temperature
with the tilt angle $\phi_2$ in [100] GBJs. The sign of the critical
current $I_c$ is positive for curves a), negative for curves c).}
\end{figure}

One of the consequences is that the Andreev level spectra are
strongly modified. Indeed, in Fig.\ref{fig:specrum1} the Andreev
spectra for two different GBJ configurations are reported. In
particular we have considered, most representatively, the case of
[100] tilt-tilt mirror ($\alpha_L=-\alpha_R=\alpha$) GBJ, with
$\phi_1=0$, $\phi_2=\pi/4$, and $\alpha=\pi/8 $ (dashed line) and
the analogous mirror [001] junction (solid line) both in the clean
limit ($Z_0=0$). As is well known, in mirror junctions,  for a given
symmetric rotation $\alpha$ ($0<\alpha<45^\circ$) around the c-axes,
depending on the quasiparticle incidence angle
$\theta$($-\pi/2<\theta<\pi/2$), two kinds of bound levels exist
\cite{Ilichev,Testa}: midgap-like states, for
$\pm\pi/4-|\alpha|<\theta<\pm\pi/4+|\alpha|$ and edgegap-like states
in the complementary intervals. The formation of midgap states, i.e.
the formation of zero energy states when the phase difference across
the junction is zero, is characteristic of d-wave Josephson junction
\cite{Hu}. Both kinds of levels determine the magnitude of the
Josephson current as well as the dependence of the maximum Josephson
current on the temperature \cite{tanaka}. The two kinds of Andreev
levels, selected for two illustrative $\theta$ values, have been
derived according to Eq. \ref{eq:energy2}. In particular, when
the levels are degenerate in energy, i.e. at $\phi=\pi$, an
asymmetric tilt of [100] junctions ($\phi_1\neq\phi_2$) opens up an
energy gap  ($E_{gap}$) in both kinds of Andreev level spectra
similar to what happens in the presence of impurities. For selected
trajectories ($\theta=0$ and $\theta=\pi/4$), the entity of this
bandgap may be simply expressed as $E_{gap}/\Delta_0 = 2
{\sqrt{1-T_{eff}^0}}|{\sin{2 \alpha}}|$ for edge-like energy states
and $E_{gap}/\Delta_0 = 2 {\sqrt{T_{eff}^0}}|{\cos{2 \alpha}}|$ for
midgap-like states, respectively.

The Josephson current may be derived directly from the energy spectrum. As a consequence it
is sensitive to the modifications derived so far. In the short
junction limit, for which $a+b<<\xi_0$, the discrete Andreev levels
determine all the Josephson current through the expression
\cite{Lofwander, Riedel}
\begin{equation}\label{eq:jos}
I_x(\varphi,\phi_1,\phi_2)=\frac{2e}{\hbar}\frac{k_FL_y}{2 \pi}\sum_n\int_{-\pi/2}^{\pi/2}
d\theta \cos(\theta)\frac{dE_n(\theta,\varphi)}{d \varphi}f(E_n(\theta))
\end{equation}
where the index $n$ labels the bound Andreev energy levels and
$f(E_n,\theta)$ is the Fermi function. As a matter of fact it turns out
that the Josephson current shows a dependence on the angles $\phi_1$ and $\phi_2$ in [100] geometries.
 Fig. \ref{fig:josephson}
shows a comparison of the dependence on the temperature of the
maximum Josephson current between asymmetric [100] and [001]  mirror
junctions. The current is normalized with respect to $I_0=2ek_FL_y
\Delta_0/h$, which is the zero temperature maximum Josephson current
through an s-wave SNS junction in the clean limit \cite{Riedel}. The
superconductive gaps in the two electrodes are assumed to obey a
Bardeen-Cooper-Schrieffer (BCS) temperature dependence. It is well
known that in this kind of junction the interplay between midgap and
edgegap states may lead to a temperature sign inversion of the
maximum Josephson current (Fig. \ref{fig:josephson} b)), i.e. to a
temperature dependent $0-\pi$ crossover \cite{tanaka,Testa} (arrow
in Fig. \ref{fig:josephson}). For the chosen value of the barrier
strength parameter $Z_0=0.7$, the modifications for increasing
values of the tilt angle $\phi_2$ are evident, showing a shift
towards higher temperatures of the $0-\pi$ crossover as $\phi_2$
increases from $0^\circ$ to $45^\circ$, see curves b) in Fig.
\ref{fig:josephson}.

In this Letter we have presented a theoretical model for the
description of electrical transport in YBCO [100] tilt valley type GBJs based on the
quasiclassical BdG equations. We have compared our results with the
well studied [001] tilt case. We have derived
 the spectral relation
in the case of [100] tilt-tilt junctions. The Andreev bound
states calculated through this equation show modifications analogous
to those caused by the presence of an insulating layer. However  the nature
of this modifications is different: it depends on the geometrical mismatch between
conducting planes.  We have evaluated the temperature
dependence of the maximum Josephson current in the case of [100]
mirror junctions and found a dependence of the
$0$-to-$\pi$ crossover temperature on the inter-electrode
misorientation angles.

The financial contribution of EU NMP.2011.2.2-6 IRONSEA project nr. 283141
is gratefully acknowledged. This work has been partially supported by EU STREP
project MIDAS, Macroscopic Interference Devices for
Atomic and Solid State Physics: Quantum Control of Supercurrents.
We wish to thank Mario Cuoco, Paola Gentile, Canio Noce for helpful discussions on this topic.

\end{document}